\documentclass[twocolumn,nofootinbib,amsmath,amssymb,amsfonts,aps,prd,balancelastpage,superscriptaddress]{revtex4-2}

\usepackage{color}
\usepackage{mathtools}
\usepackage[active]{srcltx}
\usepackage{amsthm,amstext,amscd,srcltx}
\usepackage{epsfig,graphicx,bm}
\usepackage{epstopdf, epsf}
\usepackage{dcolumn}
\usepackage{tensor}
\usepackage{graphicx,caption,subcaption}
\usepackage[colorlinks,
           linkcolor = blue,		
           citecolor = blue,
           ]{hyperref}

\newcommand{\be}{\begin{equation}}
\newcommand{\ee}{\end{equation}}

\newcommand{\bse}{\begin{subequations}}
\newcommand{\ese}{\end{subequations}}
\newcommand{\bea}{\begin{eqnarray}}
\newcommand{\eea}{\end{eqnarray}}
\newcommand{\ba}{\begin{array}}
\newcommand{\ea}{\end{array}}
\newcommand{\bc}{\begin{center}}
\newcommand{\ec}{\end{center}}

\begin{document}
\preprint{IPM/P-2012/009}  
\vspace*{3mm}

\title{Gauss--Bonnet-induced symmetry breaking/restoration during inflation}

\author{Yermek Aldabergenov}
\email{ayermek@fudan.edu.cn}
\affiliation{Department of Physics, Fudan University, 220 Handan Road, Shanghai 200433, China}

\author{Daulet Berkimbayev}
\affiliation{Department of Theoretical and Nuclear Physics, Al-Farabi Kazakh National University, 71 Al-Farabi Ave., Almaty 050040, Kazakhstan}

\begin{abstract}
\noindent

We propose a mechanism of symmetry breaking or restoration that can occur in the middle of inflation due to the coupling of the Gauss--Bonnet term to a charged scalar. The Gauss--Bonnet coupling results in an inflaton-dependent effective squared mass of the charged scalar, which can change its sign (around the symmetric point) during inflation. This can lead to spontaneous breaking of the symmetry, or to its restoration, if it is initially broken. We show the conditions, under which the backreaction of the Gauss--Bonnet coupling on the inflationary background is negligible, such that the predictions of a given inflationary model are unaffected by the symmetry breaking/restoration process.

\end{abstract}

\maketitle

\section{Introduction}

When it comes to higher curvature corrections to general relativity, the Gauss--Bonnet (GB) quadratic invariant,
\begin{equation}
    R^2_{\rm GB}\equiv R^2-4R_{\mu\nu}R^{\mu\nu}+R_{\mu\nu\rho\sigma}R^{\mu\nu\rho\sigma}~,
\end{equation}
stands out because it can avoid Ostrogradsky ghost and other instabilities, while allowing for new cosmological dynamics compared to general relativity (see \cite{Fernandes:2022zrq} for a review). On its own, it is a topological term, but when coupled to a scalar field for example, it modifies equations of motion of both background (homogeneous) fields, and scalar and tensor perturbations \cite{Cartier:2001is,Guo:2006ct,Satoh:2008ck,Guo:2009uk,Guo:2010jr} (unlike the other well-known topological quantity -- gravitational Chern--Simons term, which does not affect background dynamics and linearized scalar perturbations, when coupled to a scalar field). Inflationary effects of the GB-inflaton coupling on CMB have been extensively studied in the literature, see e.g. \cite{Jiang:2013gza,Koh:2014bka,Yi:2018gse,Odintsov:2018zhw,Odintsov:2019clh,Pozdeeva:2020apf,Pozdeeva:2024ihc} in addition to the references above.

From top-down approach, GB term is known to appear in string theory effective actions where it couples to the dilaton and moduli fields \cite{Antoniadis:1993jc,Kawai:1998ab,Kawai:1999pw}. From bottom-up point of view, it is predicted by effective field theory of inflation \cite{Weinberg:2008hq}. Most interesting cosmological scenarios can arise if the GB-scalar term is allowed to be comparable or larger than the Einstein--Hilbert term, for example, during inflation.~\footnote{It should be mentioned that in such a scenario, the suppression of cubic and higher-order curvature terms is not automatic, and should be explained by underlying UV physics or symmetry arguments.} In this case, the Gauss--Bonnet coupling can lead to new de Sitter (dS) stationary points \cite{Nojiri:2005vv,Calcagni:2005im,Nojiri:2006je,Tsujikawa:2006ph,Koivisto:2006ai,Neupane:2006dp,Granda:2017oku,Chatzarakis:2019fbn,Pozdeeva:2019agu,Vernov:2021hxo,Kawai:2021bye,MohseniSadjadi:2023amn,Pinto:2024dnm}, which can be relevant for inflation and dark energy, and which will be the main focus of this work.

As we are going to show, these GB-induced stationary points can break or restore symmetries. This mechanism provides an alternative way to realize phase transitions during (and after) inflation, in addition to those triggered by direct inflaton-scalar couplings (as in hybrid inflation \cite{Linde:1993cn,Garcia-Bellido:1996mdl,Garcia-Bellido:1996oso}), non-minimal couplings to the scalar curvature \cite{Koh:2010kg,Bettoni:2018utf,Bettoni:2018pbl,Bettoni:2019dcw,Bettoni:2021zhq,Arapoglu:2022vbf,Kierkla:2023uzo,Laverda:2023uqv,MohseniSadjadi:2023cjd,Aldabergenov:2024fws,Bettoni:2024ixe}, and thermal effects (e.g. during warm inflation \cite{Berera:1995ie,Rosa:2021gbe}, or after inflation). These phase transitions can produce and destroy cosmic defects, as well as generate gravitational waves (GW) -- see, for example, Refs. \cite{An:2022toi} and \cite{Goolsby-Cole:2017hod}.

Since physics beyond the Standard Model, such as superstrings, supergravity, and various Grand Unified Theories, often involve additional ``hidden" symmetries, it is a reasonable expectation that phase transitions can take place around the inflationary scale, leaving imprints on CMB or GW background \cite{Bettoni:2018pbl,An:2022cce,Barir:2022kzo,An:2023jxf,Kierkla:2023uzo,Bettoni:2024ixe,An:2024oui}. Furthermore, even the Standard Model symmetries can be affected during inflation or reheating stage, for example if the Higgs field gains a dynamical VEV of Hubble scale \cite{Garcia-Bellido:2001dqy,Pearce:2015nga,Yang:2015ida,Graham:2015cka,He:2018mgb,Kumar:2017ecc,Wu:2018lmx,Opferkuch:2019zbd,Wu:2019ohx}. Such models can generate lepton number during reheating once the Higgs field relaxes from its Hubble scale VEV, and can explain matter-antimatter asymmetry. Therefore, it is important to study possible mechanisms that can trigger these phase transitions.

The structure of the paper is as follows. In Section \ref{Sec_dS} we will consider the case of dS background, where the ``stationary points" are truly stationary. In Section \ref{Sec_inf} we consider an inflationary background, where the ``stationary points" are slowly evolving due to the slow-roll. This evolution can lead to the transition from symmetric to broken phase and vice versa. We derive analytical slow-roll solutions and compare them with numerical results. Section \ref{Sec_concl} is reserved for final comments and conclusion.

\section{De Sitter background}\label{Sec_dS}

Let us start with a Lagrangian for a scalar field $\phi$, which is invariant under some symmetry. For simplicity we take this symmetry to be $Z_2$, such that the model is invariant under $\phi\rightarrow -\phi$ (and $\phi$ is real). Adding the gravitational sector and the GB coupling, the Lagrangian reads,
\begin{equation}\label{L_master}
\sqrt{-g}^{\,-1}\mathcal{L} = \tfrac{1}{2}R-\tfrac{1}{2}(\partial \phi)^2-\tfrac{1}{8}\xi(\phi) R_{GB}^{2}-V(\phi)
~,
\end{equation}
where $\xi(\phi)$ is the GB coupling function. We choose the $(-,+,+,+)$ metric signature, and set $M_P=1$ unless otherwise stated.

The corresponding background equations of motion (EOM) in FLRW metric are
\begin{align}\label{EOM_phi}
    \ddot{\phi}+3H\dot{\phi}+V_{,\phi}+3\xi_{,\phi} H^2(H^2+\dot{H}) &=0~,\\
    \label{EOM_Hdot}
    2\dot H(1-\dot\xi H)-\ddot\xi H^2+\dot\xi H^3+\dot\phi^2 &=0~,\\
    3H^2(1-\dot\xi H)-\tfrac{1}{2}\dot\phi^{2}-V &=0~.\label{EOM_H}
\end{align}

Stationary points of Eq. \eqref{EOM_phi} are found from:
\begin{equation}\label{Stat_pt_eq}
    V_{,\phi}+3\xi_{,\phi} H^4=0~,
\end{equation}
where $H^2=V/3$. In Minkowski vacuum ($V=0$), the GB contribution vanishes because $H=0$. Therefore, the existence of Gauss--Bonnet-induced stationary points requires de Sitter vacuum. From \eqref{Stat_pt_eq}, the effective potential can be introduced as
\begin{equation}\label{V_eff}
    V_{\rm eff}\equiv V+\tfrac{1}{3}\int d\phi\xi_{,\phi}V^2~,
\end{equation}
where we used $H^2=V/3$. Once $\xi$ and $V$ are given, one can integrate the second term in \eqref{V_eff} and get the explicit form of the effective potential.

The simplest case is that of a constant (positive) potential $V=V_0$, where the effective potential \eqref{V_eff} becomes
\begin{equation}
    V_{\rm eff}=V_0+\tfrac{1}{3}V_0^2\xi~.
\end{equation}
When the potential is independent of $\phi$, the model has a continuous shift symmetry, $\phi\rightarrow \phi+c$ (with $c\in\mathbb{R}$), in the absence of the GB coupling, or if the GB coupling is linear in $\phi$. As $\phi$ takes an arbitrary vacuum expectation value (VEV), the shift symmetry is spontaneously broken. In this work we consider a $Z_2$ symmetry, so the GB coupling is assumed to be a function of $\phi^2$.

Our goal here is to derive the conditions under which a symmetric (symmetry breaking) vacuum of the general potential $V(\phi)$ can be overcome by the GB contribution to the effective potential \eqref{V_eff}, such that the effective vacuum becomes symmetry breaking (symmetric). 

The first condition is related to the effective mass squared around $\phi=0$,~\footnote{Since $V$ and $\xi$ are functions of $\phi^2$, their first derivatives vanish at $\phi=0$, and we get Eq. \eqref{V_eff_mass} from the effective potential \eqref{V_eff}.}
\begin{equation}\label{V_eff_mass}
    V_{{\rm eff},\phi\phi}(0)=V_{,\phi\phi}(0)+\tfrac{1}{3}\xi_{,\phi\phi}(0)V^2(0)~.
\end{equation}
Namely, we require that the GB contribution overcomes the potential contribution in magnitude,
\begin{equation}\label{Eff_mass_0_cond}
    |\xi_{,\phi\phi}(0)V^2(0)|>3|V_{,\phi\phi}(0)|~,
\end{equation}
while having the opposite sign. The second condition is the existence of a stable de Sitter minimum of $V_{\rm eff}$.

For concreteness, we will consider the scalar potential and the GB function having up to $\phi^4$ terms,
\begin{align}
    V &= V_0+\tfrac{1}{2}m^2\phi^2+\tfrac{1}{4}\lambda\phi^4~,\label{V_phi4}\\
    \xi &= \tfrac{1}{2}\alpha\phi^2+\tfrac{1}{4}\beta\phi^4~,\label{xi_phi4}
\end{align}
where $V_0,m^2,\lambda,\alpha,\beta$ are real constants ($\alpha$ with mass dimension $-4$, and $\beta$ with mass dimension $-6$), and we assume $V_0,\lambda>0$. The constant term in $\xi$ is irrelevant, as it corresponds to the topological term in the action.

Next, we will derive stable dS stationary points of $V_{\rm eff}$ that spontaneously break or restore the $Z_2$ symmetry.

\subsection{GB-induced SSB}\label{sec_GB_SSB}

For GB-induced spontaneous symmetry breaking (SSB), we start from the potential \eqref{V_phi4} with $m^2>0$, such that the symmetry is unbroken in the absence of the GB coupling. We first destabilize the symmetric point by imposing (from \eqref{Eff_mass_0_cond})
\begin{equation}\label{alpha_cond_0}
    |\alpha|>3m^2/V_0^2~,
\end{equation}
as well as $\alpha<0$ to counteract the positive $m^2$ from the potential.

The next step is to find a stable symmetry breaking minimum. By using \eqref{V_phi4} and \eqref{xi_phi4}, we can get an explicit form of the effective potential \eqref{V_eff}, and study its stationary points satisfying
\begin{align}
\begin{aligned}\label{dV_eff_phi4}
    V_{{\rm eff},\phi} &=\phi\Big[m^2+\lambda\phi^2\\
    &+\tfrac{1}{3}(\alpha+\beta\phi^2)(V_0+\tfrac{1}{2}m^2\phi^2+\tfrac{1}{4}\lambda\phi^4)^2\Big]=0~.
\end{aligned}
\end{align}
It is necessary to have non-zero quartic (or higher-order) terms in $\xi$ and/or $V$ for the existence of non-zero stationary points. If the quartic terms are absent, that is, $\lambda=\beta=0$, Eq. \eqref{dV_eff_phi4} can be written as (using $\alpha=-|\alpha|$)
\begin{equation}\label{phi_no_quartic}
    \phi^2=\frac{2}{m^2}\Big(\sqrt{\frac{3m^2}{|\alpha|}}-V_0\Big)~.
\end{equation}
By using Eq. \eqref{alpha_cond_0}, the right side of \eqref{phi_no_quartic} becomes negative, i.e. no stationary points with $\phi\neq 0$ exist.

In the presence of the quartic terms, Eq. \eqref{dV_eff_phi4} is a fifth-order polynomial in $\phi^2$. In certain situations we can solve \eqref{dV_eff_phi4} perturbatively, such as when $V_0$ is much larger than $m^2\phi^2$ and $\lambda\phi^4$. In this case, the solution to \eqref{dV_eff_phi4} is
\begin{equation}\label{VEV_GB_SSB}
    \langle\phi\rangle^2_{\rm eff}\equiv\nu^2_{\rm eff}\simeq\frac{V_0^2|\alpha|-3m^2}{V_0^2\beta+3\lambda}~.
\end{equation}
This approximation can be applied, for example, to a scalar field in the inflationary background, $V_0\simeq 3H^2_{\rm inf}$, if $m$ and $\nu_{\rm eff}$ is not too large compared to the Hubble function $H_{\rm inf}$ (by order of magnitude).

GB-induced SSB was proposed in \cite{Liang:2019fkj} as a novel baryogenesis mechanism, where the symmetry breaking was studied in a cosmological background with the equation of state, $w\neq -1$ (see also \cite{MohseniSadjadi:2024ejb} for GB-induced symmetry restoration in the context of quintessence dark energy). In contrast, here we consider an inflationary background, and study the general conditions for GB-induced symmetry breaking and restoration.

\subsection{GB-induced symmetry restoration}

Next, we discuss the opposite situation where the mass term of the scalar potential is tachyonic at the symmetric point, $m^2<0$, and the GB coupling is used to stabilize the scalar at this point, and restore the symmetry. For this, it suffices to consider quadratic GB coupling, and we parametrize $V$ and $\xi$ as
\begin{equation}\label{GB_SR_potentials}
    V=V_0-\tfrac{1}{2}\lambda\nu^2\phi^2+\tfrac{1}{4}\lambda\phi^4~,~~~\xi=\tfrac{1}{2}\mu^{-4}\phi^2~,
\end{equation}
where we replaced $m^2\rightarrow -\lambda\nu^2$ (such that $\nu$ is the VEV of $\phi$ in the absence of the GB coupling) and $\alpha\rightarrow\mu^{-4}$, where $\mu$ is a real constant with mass dimension one.

First, let us look into how the GB coupling affects the VEV of the scalar. Stationary points of the effective potential are found from
\begin{align}
\begin{aligned}\label{dVeff_restoration}
    V_{{\rm eff},\phi} &=\lambda\nu^2\phi\bigg[\frac{\Omega}{3}Y^2+\Big(\frac{\phi^2}{\nu^2}-1\Big)\\
    &+\frac{\Omega}{6}Y\Big(\frac{\phi^2}{\nu^2}-1\Big)^2+\frac{\Omega}{48}\Big(\frac{\phi^2}{\nu^2}-1\Big)^4\bigg]=0~,
\end{aligned}
\end{align}
where we use the shorthand notation
\begin{equation}\label{Omega_Y_defs}
    \Omega\equiv\frac{\lambda\nu^6}{\mu^4}~,~~~Y\equiv\frac{V_0}{\lambda\nu^4}-\frac{1}{4}~.
\end{equation}
Both $\Omega$ and $Y$ are positive-definite ($Y$ is positive-definite because otherwise, the symmetry breaking vacuum becomes AdS or Minkowski, contradicting our initial assumption of a dS vacuum). This implies that out of the four terms in the square brackets of \eqref{dVeff_restoration}, only the second term, $(\phi^2/\nu^2-1)$, can be negative. Therefore, if there is a non-zero stationary point of the effective potential, it is necessarily smaller than $\nu$ in magnitude.

We can again search for a perturbative solution to \eqref{dVeff_restoration}. By comparing the first, third, and fourth terms (all of which are positive), it is clear that if $Y\sim\mathcal{O}(1)$ or larger, then the first term is dominant, and we can approximate the stationary point as
\begin{equation}\label{GB_VEV_restoration}
    \frac{\phi^2}{\nu^2}\approx 1-\frac{\Omega}{3}Y^2=1-\frac{(V_0-\lambda\nu^4/4)^2}{3\lambda\nu^2\mu^4}~,
\end{equation}
by ignoring the third and fourth terms. In the limit $\mu\rightarrow\infty$, we recover the original VEV, $\phi=\pm\nu$.

As for the effective mass of $\phi$, we have
\begin{equation}\label{m_eff_restoration}
    V_{{\rm eff},\phi\phi}(0)=\lambda\nu^2\big[\tfrac{\Omega}{48}(1+4Y)^2-1\big]~.
\end{equation}
Using the definitions \eqref{Omega_Y_defs} of $\Omega$ and $Y$, we find that the effective mass \eqref{m_eff_restoration} is positive, zero, or negative, if (respectively)
\begin{equation}\label{mu_SSB}
    \mu^4<\frac{V_0^2}{3\lambda\nu^2}~,~~~\mu^4=\frac{V_0^2}{3\lambda\nu^2}~,~~~\mu^4>\frac{V_0^2}{3\lambda\nu^2}~.
\end{equation}
The first case describes GB-induced symmetry restoration, while the third case is the symmetry breaking scenario modified by the GB-scalar coupling, where the scalar (effective) VEV is given by \eqref{GB_VEV_restoration} for $Y\geq\mathcal{O}(1)$. In the second case, the scalar is effectively massless.

\section{Application to inflation}\label{Sec_inf}

During inflation we have a quasi-dS background with Hubble function, $H_{\rm inf}^2\simeq V_{\rm inf}/3$,~\footnote{In the previous section, dS expansion was driven by a cosmological consant which we denote as $V_0$. In this section, we consider an inflationary quasi-dS background, and therefore replace $V_0$ by an inflationary potential $V_{\rm inf}$. A small positive cosmological constant (dark energy) can be included as well, but it is insignificant during inflation, so we ignore it here.} slowly changing (decreasing) with time. That is, we can introduce slow-roll parameters
\begin{equation}
    \epsilon\equiv -\dot H/H^2~,~~~\eta\equiv\dot\epsilon/(H\epsilon)~,
\end{equation}
such that $\epsilon,|\eta|\ll 1$ during inflation. We assume inflation is driven, as usual, by a real scalar field (inflaton) $\chi$ with a suitable potential $V_{\rm inf}(\chi)$. The results of the previous section can be applied by replacing $V_0\rightarrow V_{\rm\inf}(\chi)$.

For simplicity we ignore the GB coupling of the inflaton, and extend the EOM \eqref{EOM_phi}--\eqref{EOM_H} as
\begin{align}
    \ddot{\chi}+3H\dot{\chi}+V_{,\chi} &=0~,\label{EOM_inf_chi}\\
    \ddot{\phi}+3H\dot{\phi}+V_{,\phi}+3\xi_{,\phi} H^2(H^2+\dot{H}) &=0~,\label{EOM_inf_phi}\\
    2\dot H(1-\dot\xi H)-\ddot\xi H^2+\dot\xi H^3+\dot\phi^2+\dot\chi^2 &=0~,\label{EOM_inf_Hdot}\\
    3H^2(1-\dot\xi H)-\tfrac{1}{2}(\dot\phi^{2}+\dot\chi^2)-V &=0~,\label{EOM_inf_H}
\end{align}
where $V$ is the total scalar potential which we parametrize as
\begin{equation}\label{V_inf_quartic}
    V=V_{\rm inf}+\tfrac{1}{2}m^2\phi^2+\tfrac{1}{4}\lambda\phi^4~.
\end{equation}
For the inflaton potential we will use a simple Starobinsky/$\alpha$-attractor potential
\begin{equation}\label{V_inf_Star}
    V_{\rm inf}=\tfrac{1}{2}M^2(1-e^{-\chi})^2~,
\end{equation}
although one can consider any single-field inflationary model of choice, satisfying CMB data \cite{Planck:2018jri}.

\subsection{Symmetry restoration scenario}

Consider a theory with symmetry-preserving scalar potential, for instance, given by \eqref{V_inf_quartic} with $m^2>0$. We now spontaneously break the symmetry by turning on the GB-coupling \eqref{xi_phi4} with large enough negative $\alpha$ such that the effective mass-squared of $\phi$ becomes negative (as described in Section \ref{sec_GB_SSB}). However, since $V_{\rm inf}$ decreases with time during inflation, the symmetry can eventually be restored, even before the end of inflation. This can be seen from the effective VEV (from \eqref{VEV_GB_SSB})
\begin{equation}\label{SR_inf_nu}
    \nu^2_{\rm eff}\simeq\frac{V_{\rm inf}^2|\alpha|-3m^2}{V_{\rm inf}^2\beta+3\lambda}~,
\end{equation}
where we assume that $\beta>0$, and/or subdominant to the $\lambda$ term, such that the existence of this VEV requires $V_{\rm inf}^2>3m^2/|\alpha|$.

Suppose that at some point in time $t_1$ (e.g. when CMB scales exit the horizon), the condition $V_{\rm inf}^2(t_1)>3m^2/|\alpha|$ holds. Eventually we reach the equality $V_{\rm inf}^2(t_{\rm c})=3m^2/|\alpha|$ at some later time $t_{\rm c}$ (critical time), signalling the restoration of the symmetry, since $\nu_{\rm eff}$ becomes zero at this point and afterwards. If $t_1<t_{\rm c}<t_2$, where $t_2$ denotes the end of inflation, this scenario results in symmetry restoration in the middle of observable inflation. Alternatively, the symmetry can be restored (with the help of the Gauss--Bonnet coupling) after inflation, as was considered in \cite{Liang:2019fkj}, where $U(1)_{B-L}$ is initially broken by GB-coupling, and is restored after inflation (and after Affleck--Dine baryogenesis).

\textbf{Slow-roll solution}. Under certain assumptions we can obtain analytical slow-roll solution. It is convenient to change time variable from $t$ to the number of efolds $N$ growing with time in our convention ($\dot N=H$). Scalar equations read
\begin{align}
    \chi''+(3-\epsilon)\chi'+V_{,\chi}H^{-2} &=0~,\label{EOM_inf_chi_N}\\
    \phi''+(3-\epsilon)\phi'+V_{,\phi}H^{-2}+3\xi_{,\phi} H^2(1-\epsilon) &=0~,\label{EOM_inf_phi_N}
\end{align}
where $'\equiv d/dN$. Although the constraint equation \eqref{EOM_inf_H} is cubic in $H$, when expressed in terms of the efold time, it becomes a quadratic equation for $H^2$,
\begin{equation}
    3(1-\xi'H^2)H^2-\tfrac{1}{2}H^2(\chi'^2+\phi'^2)-V=0~,
\end{equation}
whose solution can be written as
\begin{equation}
    H^2=\frac{2V}{3-\tfrac{1}{2}(\chi'^2+\phi'^2)}\bigg\{1+\sqrt{1-\frac{4\xi'V}{3[1-\tfrac{1}{6}(\chi'^2+\phi'^2)]^2}}\bigg\}^{-1}~.
\end{equation}
This form of the solution (as opposed to the standard quadratic formula) avoids catastrophic cancellation in the limit $\xi'\rightarrow 0$.

In order to realize slow-roll inflation, in addition to the usual slow-roll conditions, $\epsilon,|\eta|\ll 1$, we also introduce the GB slow-roll conditions $|\omega|,|\omega'|\ll 1$, where $\omega\equiv\xi'H^2$. Furthermore, if we want to minimize the backreaction of the GB coupling on the $\chi$-driven slow-roll, we can demand
\begin{equation}\label{GBSR_cond}
    |\omega|,|\omega'|\ll \epsilon\ll 1~.
\end{equation}
The first inequality is not a necessary condition for our mechanism to work, but it will help to keep the predictions of the Starobinsky model intact, and will allow us to obtain analytical results.

Here we fix the potential $V$ as in Eqs. \eqref{V_inf_quartic} and \eqref{V_inf_Star}, and the GB coupling as
\begin{equation}\label{xi_alpha_beta}
    \xi=-\tfrac{1}{2}|\alpha|\phi^2+\tfrac{1}{4}\beta\phi^4~.
\end{equation}

Under the conditions \eqref{GBSR_cond}, we can expect the usual slow-roll approximation to hold for the inflaton $\chi$. We set $\phi=\phi'=\phi''=0$ (as will be justified below), and derive the slow-roll solution for the inflaton. By introducing $y\equiv e^{-\chi}$, the slow-roll solution to \eqref{EOM_inf_chi_N} can be written as
\begin{equation}
    y\simeq\frac{1}{2(C-N)}~,
\end{equation}
where $C$ is the integration constant, and we used $\epsilon\rightarrow 0$, $H^2\simeq V/3$, and ignored $\chi''$. We can eliminate $C$ by introducing $y_*$ as the value of $y$ at the horizon exit, and setting $N_*=0$. Then $y_*=1/(2C)$, and the solution takes the form
\begin{equation}\label{y_sol}
    y\simeq\frac{y_*}{1-2y_*N}~\Rightarrow~~N\simeq\frac{1}{2}\Big(\frac{1}{y_*}-\frac{1}{y}\Big)~,
\end{equation}
where we inverted the solution to obtain $N(y)$. This can be used to obtain the number of efolds from the horizon exit until the end of inflation, $\Delta N=N_{\rm end}-N_*=N_{\rm end}$. In the Starobinsky model, the inflaton value at the horizon exit satisfies $y_*=e^{-\chi_*}\ll 1$, as well as $y_*\ll y_{\rm end}$. Therefore we can write $\Delta N\simeq 1/(2y_*)$.

Scalar spectral tilt $n_s$ and tensor-to-scalar ratio $r$ can be written with the help of the potential slow-roll parameters $\epsilon_V\equiv V^2_\chi/(2V^2)$ and $\eta_V\equiv V_{\chi\chi}/V$ as
\begin{equation}
    n_s\simeq 1+2\eta_V-6\epsilon_V~,~~~r\simeq 16\epsilon_V~.
\end{equation}
At the leading order in $y_*$, and after substituting $y_*\simeq 1/(2\Delta N)$, we get the well known result
\begin{equation}
    n_s\simeq 1-2/\Delta N~,~~~r\simeq 8/\Delta N^2~,
\end{equation}
which is in a good agreement with CMB data \cite{Planck:2018jri} for around $55$ efolds of inflation. Lastly, the parameter $M$ is fixed by matching the CMB amplitude of scalar perturbations,
\begin{equation}
    A_s\approx\frac{V}{24\pi^2\epsilon_V}\Big|_{\chi_*}\approx\frac{\Delta N^2 M^2}{24\pi^2}~,
\end{equation}
with the observed value $A_s\approx 2.1\times 10^{-9}$.
For $\Delta N=55$, we find $M\approx 1.28\times 10^{-5}$.

Having obtained the analytical slow-roll solution, it is straightforward to introduce initial symmetry breaking in $\phi$-direction, which is then restored during inflation. This can be done without affecting the predictions of the Starobinsky model, if the quadratic and quartic $\phi$-terms in Eq. \eqref{V_inf_quartic} are much smaller than $V_{\rm inf}$ during inflation. In particular, we assume that around the horizon exit, the symmetry is broken, $\phi=\nu_{\rm eff}$, and the effective VEV of $\phi$ is given by \eqref{SR_inf_nu} (this also requires slow-roll conditions to hold, since \eqref{SR_inf_nu} is obtained for quasi-dS background).

For simplicity, we set $\beta=0$, while $\alpha$ can be fixed by our choice of the critical time $t_c$, or $N_c$, when the symmetry is restored. From \eqref{y_sol} we then find $y_c=y_*/(1-2y_*N_c)$, and insert it into the condition for vanishing effective $\phi$-mass (see the equality in Eq. \eqref{mu_SSB}),
\begin{equation}\label{alpha_critical}
    V_{\rm inf}^2(y_c)=\tfrac{1}{4}M^4(1-y_c)^4=3m^2/|\alpha|~,
\end{equation}
to find $|\alpha|$. Since $|\alpha|$ has dimensions of mass$^{-4}$, we can write $|\alpha|\equiv 1/\mu^4$ and work with the mass parameter $\mu$. 

To be more specific with the parameters, let us choose $N_c=30$, that is, the symmetry is restored around $30$ efolds after the horizon exit of the CMB scale. If the observable inflation lasts for $\Delta N=55$ (i.e. $y_*\approx 1/110$), this leads to $y_c=1/50$, and from \eqref{alpha_critical} we have
\begin{equation}\label{mu_critical}
    \mu=\frac{49M}{12^{1/4}50\sqrt{m}}~.
\end{equation}
Other possible values of $N_c$ are $0<N_c<55$ for the symmetry restoration during observable inflation. In our conventions, $N_c<0$ would mean symmetry restoration before the observable inflation, while $N_c>55$ -- after inflation.

In our examples we fix $\lambda=0.1$ (for perturbativity in the scalar potential we need $\lambda\ll 1$), so that the only free parameter left is $m$. Since the GB parameter $\mu$ depends on $m$ through \eqref{mu_critical}, by demanding $|\omega|,|\omega'|\ll\epsilon$ (to avoid the GB backreaction) the parameter $m$ will be bounded from above as follows.

Since $\omega$ is proportional to $\xi_{,\phi}\propto\phi$, after the symmetry restoration $\omega$ will vanish automatically, so we only have to consider the behavior of $\omega$ at the start of observable inflation, when $\phi$ is at its effective VEV,
\begin{equation}\label{nu_eff_y}
    \nu_{\rm eff}^2\simeq \frac{1}{3\lambda}\bigg[\frac{M^4}{4\mu^4}(1-y)^4-3m^2\bigg]~,
\end{equation}
which is obtained from \eqref{SR_inf_nu} by using $\beta=0$, $|\alpha|\equiv 1/\mu^4$, and \eqref{V_inf_Star}. We now write $\omega$ as
\begin{equation}\label{omega_SR}
    \omega\equiv\xi'H^2\simeq -\frac{M^2(\phi^2)'}{12\mu^4}~,
\end{equation}
by using $H^2\simeq V/3\simeq M^2/6$. After substituting the effective VEV \eqref{nu_eff_y} for $\phi^2$ in \eqref{omega_SR}, and using inflationary slow-roll parameters, $\epsilon\simeq y'^2/(2y^2)\simeq\epsilon_V\simeq 2y^2$, we get
\begin{equation}
    \omega\simeq -\frac{M^6\epsilon_V}{36\lambda\mu^8}\simeq -2.87\times 10^{11}m^4\epsilon_V~,
\end{equation}
where at the last step we used \eqref{mu_critical} and $\lambda=0.1$. Therefore, we impose 
\begin{equation}\label{m_cond}
    m^4\ll 3.48\times 10^{-12}~,
\end{equation}
to suppress the GB slow-roll parameter according to \eqref{GBSR_cond}, which in turn will negate the backreaction of the GB coupling, and allow us to derive analytical slow-roll solutions while keeping standard predictions of single-field models (in our example, the Starobinsky model). One can also verify that $m^2\phi^2$ and $\lambda\phi^4$ terms (when $\phi=\nu_{\rm eff}$) are suppressed compared to $V_{\rm inf}$ under the condition \eqref{m_cond}.

Another important consideration is the effective mass of $\phi$ around the VEV $\nu_{\rm eff}$. Repeating the above approximation and parameter choices, we get
\begin{equation}
    V_{{\rm eff},\phi\phi}(\nu_{\rm eff})\simeq 0.09m^2-1.22\times 10^{10}m^6~.
\end{equation}
When $m$ satisfies \eqref{m_cond}, the second term is unimportant, and the effective $\phi$-mass at its VEV is close to the mass parameter $m$. This means that for small values of $m$, the kinetic energy of $\phi$ can become large enough to prevent or delay the relaxation of $\phi$ around its effective VEV. This could significantly affect the symmetry restoration time as well, forcing us to use numerical solutions. Furthermore, $m$ cannot be arbitrarily smaller than the inflaton mass, due to the isocurvature constraints at the CMB scale. We find that the values of $m$ around $10^{-4}$ allow us to derive more or less accurate analytical solution, while adhering to the upper limit \eqref{m_cond} needed to suppress the GB backreaction. If we only require the latter condition and do not care about analytical results, $m$ could be smaller, but the lower bound would be imposed by the isocurvature constraints. Essentially, with the values $m\ll H_{\rm inf}$, where $H_{\rm inf}\sim 10^{-5}$ in the Starobinsky model, there is a danger of overproduction of isocurvature modes \cite{Linde:1985yf} (although it would also depend on possible $\phi-\chi$ couplings). As for more precise isocurvature constraints on the parameters, it requires a separate analysis of the multi-field perturbations in the presence of the GB term, which is beyond the scope of this work.

\begin{figure}
\centering
  \centering
  \includegraphics[width=1\linewidth]{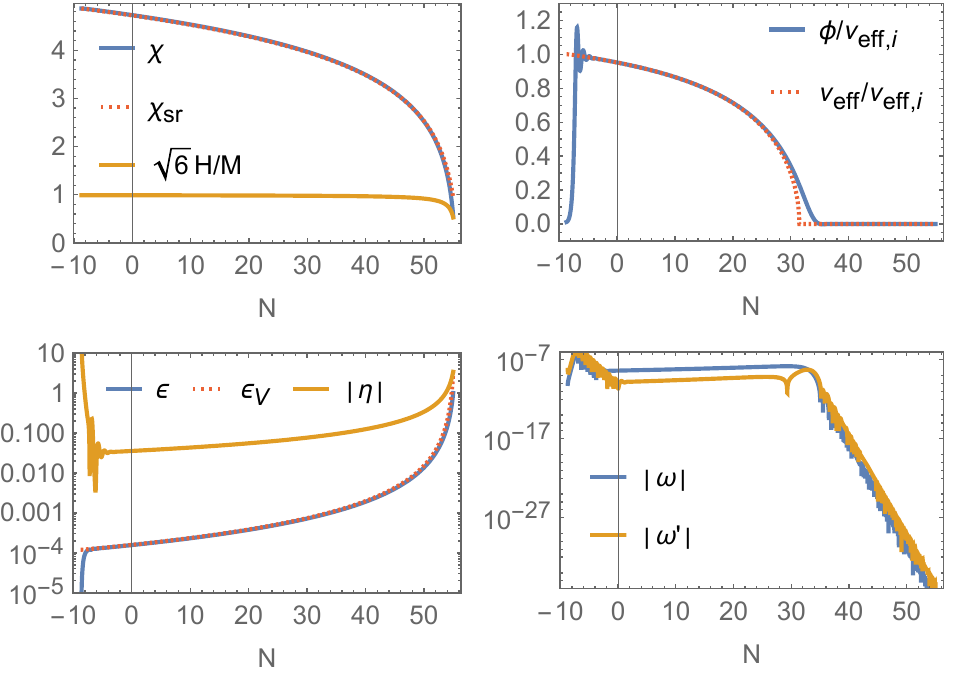}
\captionsetup{width=1\linewidth}
\caption{Numerical inflationary solution for the potential \eqref{V_inf_quartic} and \eqref{V_inf_Star}, and GB function \eqref{xi_alpha_beta} (see main text for the parameter choice). Top-left: evolution of $\chi$ and its slow-roll approximation $\chi_{\rm sr}$; orange curve shows the numerical Hubble function. Top-right: evolution of $\phi$ and its effective VEV ($\nu_{\rm eff,i}$ is the value of $\nu_{\rm eff}$ at the initial time). Bottom-left: slow-roll parameters $\epsilon$, $\eta$, and $\epsilon_V$. Bottom-right: GB slow-roll parameter $\omega$ and its velocity.}\label{Fig_sym_res}
\end{figure}

Finally, let us show the full numerical solution for $\chi$ and $\phi$, and how it compares to our analytical result. Figure \ref{Fig_sym_res} shows numerical solution of the system \eqref{EOM_inf_chi}-\eqref{EOM_inf_H} with negligible GB backreaction. The parameters used are $M=1.28\times 10^{-5}$, $\lambda=0.1$, $m=10^{-4}$, and $\mu$ is given by \eqref{mu_critical} (this leads to the effective VEV close to the inflationary scale, $\nu_{\rm eff}\approx 7.15\times 10^{-5}$). The initial conditions for the inflaton are $y=1/130$ and $\chi'=0.001$, while the initial conditions for the symmetry breaking field $\phi$ are taken as $\phi=\phi'=H/(2\pi)\simeq M/(2\sqrt{6}\pi)$, as can be expected from quantum fluctuations around the symmetric point in de Sitter background \cite{Starobinsky:1994bd}. We choose positive values for $\phi$ and $\phi'$ initial conditions, but their sign is randomly fixed among different Hubble patches, resulting in domain wall formation (or cosmic strings for a spontaneously broken $U(1)$ symmetry). The point $N=0$ corresponds to the horizon exit of the CMB reference scale, and the numerical integration starts ten efolds earlier. Plot on the top-left of Figure \ref{Fig_sym_res} shows the numerical evolution of $\chi$ compared to its analytical solution \eqref{y_sol}, as well as the rescaled Hubble function. Plot on the top-right shows the numerical evolution of $\phi$ versus the effective VEV $\nu_{\rm eff}$ given by \eqref{nu_eff_y} (both rescaled by the VEV value $\nu_{\rm eff,i}$ at the initial time $N=-10$). On the bottom-left of Figure \ref{Fig_sym_res} we can see the numerical evolution of SR parameters $\epsilon$, $\epsilon_V$ (for the inflaton potential), and $\eta$, where to a good approximation we have $\epsilon\simeq\epsilon_V$. In comparison to $\epsilon$, the GB slow-roll parameter $\omega$ and its derivative are suppressed, as can be seen on the bottom-right plot. These plots show that under appropriate parameter choice, we can effectively reproduce single-field inflation, and at the same time introduce an initially broken symmetry of the $\phi$-field, which can then be restored  during (or after) inflation. Here, for example, we chose the restoration time $N_c\approx 30$, such that $\nu_{\rm eff}=0$ onward. As is shown in Figure \ref{Fig_sym_res} (top-right), the actual restoration time can be delayed by a few efolds, since the trajectory needs some time to stabilize.

\subsection{Symmetry breaking scenario}

Here we will consider the opposite situation: suppose that on top of the symmetry-breaking potential $V$, we have a GB contribution which restores the symmetry at the beginning of inflation. The GB-induced effective mass decreases with time, and turns tachyonic at some point during inflation, leading to spontaneous breaking of the symmetry.

We take the scalar potential and GB coupling from \eqref{GB_SR_potentials}, and replace $V_0$ with 
\begin{equation}\label{V_inf_SSB}
    V_{\rm inf}=\tfrac{1}{4}\lambda\nu^4+\tfrac{1}{2}M^2(1-e^{-\chi})^2~.
\end{equation}
The reason for adding the constant term $\lambda\nu^4/4$ is to uplift the vacuum at $\chi=0$ and $\phi=\nu$ from AdS to Minkowski.

The GB parameter $\mu$ can be fixed from Eq. \eqref{mu_SSB} as
\begin{equation}
    \mu^4=\frac{V_{\rm inf}^2(\chi_c)}{3\lambda\nu^2}~,
\end{equation}
which leads to the vanishing effective mass of $\phi$ at some critical point of choice, $\chi_c$. We again set $y_c=e^{-\chi_c}=1/50$, such that $N_c\approx 30$ in the convention where $N=0$ is the horizon exit time of the CMB scale. For the numerical solution we again choose $\lambda=0.1$, and for $\nu$ we choose the value $\nu=10^{-3}$, which leads to negligible GB backreaction according to \eqref{GBSR_cond} (we will comment on other possible values of $\nu$ below). In this case, as before, the single-field solution \eqref{y_sol} can be used to describe the inflationary background on which the symmetry breaking happens, while the scalar potential is dominated by the Starobinsky potential during inflation.

\begin{figure}
\centering
  \centering
  \includegraphics[width=1\linewidth]{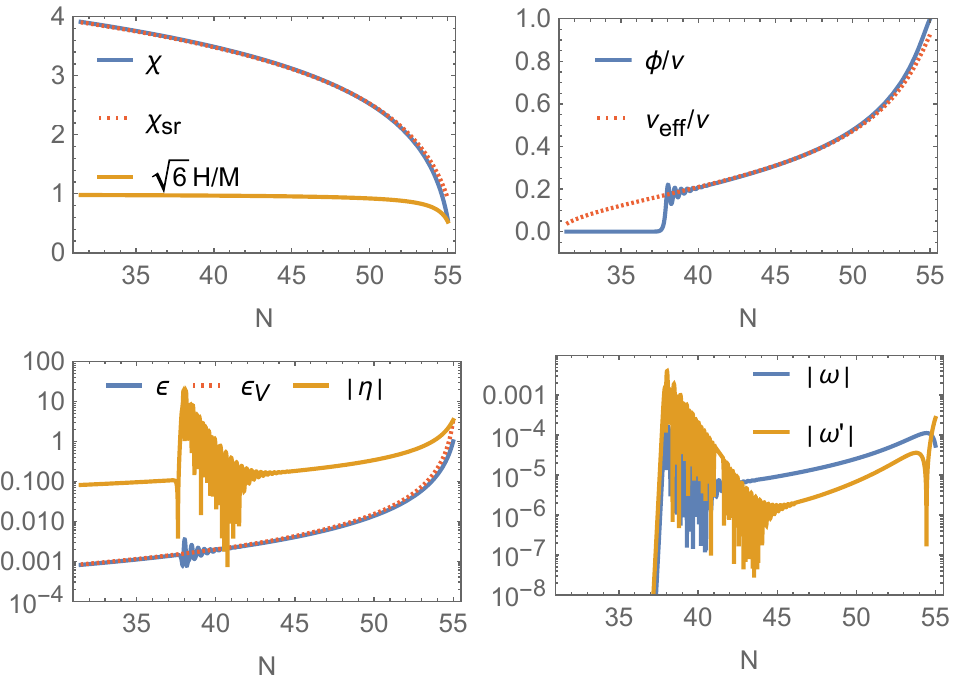}
\captionsetup{width=1\linewidth}
\caption{Numerical inflationary solution for the potentials \eqref{GB_SR_potentials} ($V_0$ is replaced with \eqref{V_inf_SSB}), during the last $\sim 25$ efolds (after the symmetry breaking). The plot description is the same as in Figure \ref{Fig_sym_res}.}\label{Fig_SSB}
\end{figure}

From the horizon exit to $N=N_c$, the scalar $\phi$ is stabilized around the symmetric point by the GB-induced mass. After $N_c$, the mass-squared becomes negative, and quantum fluctuations destabilize $\phi$ from zero. We take $\phi=\phi'\simeq M/(2\sqrt{6}\pi)$ as the initial condition for the numerical solution after the critical time $N_c\approx 30$. The solution is shown in Figure \ref{Fig_SSB}, where in the top-left plot we compare the numerical trajectory of $\chi$ with its slow-roll approximation \eqref{y_sol} during the last $\sim 25$ efolds, corresponding to the symmetry breaking phase of inflation. The top-right plot shows the trajectory of $\phi$ (in blue), following the effective VEV \eqref{GB_VEV_restoration} (with $V_0$ replaced by the inflaton potential \eqref{V_inf_SSB}). As can be seen from this plot, the $\phi$-field initially spends some time around the origin, before falling into the GB-induced vacuum $\nu_{\rm eff}$, where it quickly stabilizes after several oscillations. These damped oscillations induce transient oscillations of the slow-roll parameters ($\epsilon$, $\eta$, and $\omega$), as shown in the bottom row of Figure \ref{Fig_SSB}.

Let us comment on other possible values of $\nu$. For the values $\nu\gg 10^{-3}$, the backreaction of the GB coupling becomes large (which can impact the inflationary observables), and the approximation \eqref{GB_VEV_restoration} of the effective VEV of $\phi$ breaks down. On the opposite side, $\nu\ll 10^{-3}$, the GB backreaction is small, but due to the larger suppression of the $\phi$-potential against the inflaton potential, $\phi$ spends much more time around its origin, and the symmetry breaking time is considerably delayed compared to the analytical prediction.

\section{Conclusions}\label{Sec_concl}

In this work, we demonstrated how Gauss--Bonnet-scalar coupling can lead to the transition from a symmetric to a broken phase (and vice versa) during inflation, due to the appearance of GB-induced inflaton-dependent effective mass of the symmetry breaking scalar. By Taylor-expanding the scalar potential $V(\phi)$ and the GB coupling function $\xi(\phi)$ up to quartic terms,~\footnote{In principle, one can consider different functions for $V(\phi)$ and $\xi(\phi)$ under suitable conditions for GB-induced symmetry breaking. The condition for the absence of GB backreaction on inflation would be unchanged, and is given by \eqref{GBSR_cond} in terms of the slow-roll parameters.} we considered a simple inflationary scenario where the GB-induced effects and the dynamics of the charged scalar $\phi$ do not backreact on the inflationary slow-roll solutions (driven by a separate inflaton field), and predictions of a given single-field inflationary model are unaffected. Therefore, our mechanism can apply to any viable inflationary model, and introduce a new way to trigger phase transitions in the middle of inflation (or during reheating as was already discussed in \cite{Liang:2019fkj}). One can envision the application of the GB-induced symmetry breaking to the Standard Model Higgs field as well. For example, in Refs. \cite{Garcia-Bellido:2001dqy,Pearce:2015nga,Yang:2015ida,Opferkuch:2019zbd,Wu:2019ohx} it was shown that nontrivial evolution of the Higgs field during inflation and reheating can lead to successful leptogenesis. On the other hand, for a ``hidden" scalar with GB coupling, the implications for reheating also deserve a separate study due to the complications and uncertainties related to the presence of two scalar fields (the inflaton and the symmetry breaking scalar). For example, the inflaton can also couple to the GB term, and both scalars can interact with the Standard Model fields depending on the symmetries of a given model and charge assignment of the fields.

The ``no-backreaction" scenario, which we focused on in this paper, allows us to analytically study the inflationary solution, as well as the symmetry breaking solution for $\phi$. A comparison with numerical results was also made, as shown in Figs. \ref{Fig_sym_res} and \ref{Fig_SSB}. The absence of the GB backreaction is guaranteed by Eq. \eqref{GBSR_cond}, which also means that the scalar and tensor power spectra (unsurprisingly) receive negligible GB corrections. More specifically, the oscillations of the slow-roll parameters, which can be seen in Figs. \ref{Fig_sym_res} and \ref{Fig_SSB}, do not lead to significant deviations from the usual (single-field) inflationary power spectrum thanks to the suppression condition \eqref{GBSR_cond}. However, one can consider more general models with larger GB coupling, for instance with $\omega$ comparable or larger than $\epsilon$, which can lead to full multi-field inflation. With the inclusion of a possible GB coupling of both scalars, this warrants a more thorough investigation of multi-field GB-coupled inflation, particularly in order to analyze the evolution of multi-scalar perturbations in GB-coupled models. This could be interesting from the point of view of small-scale inflationary phenomenology, such as the enhancement of scalar and tensor perturbations at sub-CMB scales.

Some examples of non-trivial effects of large GB couplings on inflationary perturbations were discussed in Refs. \cite{Kawai:2021edk,Kawaguchi:2022nku,Ashrafzadeh:2023ndt,Solbi:2024zhl} in relation to primordial black holes. In \cite{Addazi:2024gew} it was shown that a two-field inflation, where one of the fields is coupled to the GB term, can produce sound speed resonance of GW, which can potentially be probed by future observations.~\footnote{Alternatively GW can be generated due to the GB-induced dip in the sound speed, as found in Ref. \cite{Kawai:2023nqs}.} The resonance is triggered by oscillations of the scalar field with large GB coupling (and backreaction). Although such oscillations also occur in our SSB scenario (for the charged scalar $\phi$), we have confirmed that the GB suppression \eqref{GBSR_cond} rules out the resonance effect. In summary, models that violate \eqref{GBSR_cond} (this includes possible transient slow-roll violations during inflation) and include multi-scalar GB coupling, can lead to highly non-trivial inflationary trajectories, and should be studied in more detail by numerical analysis.

\providecommand{\href}[2]{#2}\begingroup\raggedright\endgroup

\end{document}